# THE EVOLUTION OF GALAXIES

JAN PALOUŠ

*Astronomical Institute*
*Academy of Sciences of the Czech Republic*
*Boční II 1401, Praha 4 Spořilov, Czech Republic*
E-mail: `palous@ig.cas.cz`

*Abstract.* The evolution of galaxies results from a combination of internal and external processes. The star formation is an internal process transforming cold and dense cores of molecular clouds to stars. It may be triggered internally by expanding shells, or externally, e.g., by galaxy collisions. The gas accretion and galaxy merging events are external contributors to galaxy evolution. They compete with another internal process of galaxy evolution, which is the secular evolution redistributing the mass and angular momentum inside of galaxies as a consequence of bar and spiral arms formation. As a nearby example of gas accretion we mention the interacting system Milky Way − Large and Small Magellanic Clouds. Stripping of ISM in galaxy clusters is reviewed as another example of interaction of galaxies with their environment.

## 1. INTRODUCTION

Galaxies are composed of stars, interstellar matter (ISM) and dark matter (DM). While the DM dominates galaxies gravitationally, the first two components produce photons that are detected with the telescopes. As an example of a gas-rich system, we refer to the spiral galaxy NGC 1350. It shows the ISM in the disk with spiral arms reaching from the periphery to the very center of it. On the other hand, the elliptical galaxy M87 is composed mostly from stars without any gaseous disk. In spiral disk galaxies we observe the star formation changing ISM into stars. At the same time, stars return a fraction of their mass to the ISM. This gas recycling inserts the heavy elements, products of nuclear fusion in stars, into the ISM causing the chemical evolution of galaxies.

Formation and evolution of galaxies is a complex process: the mass arrives to galaxies in an accretion process or in merging events when smaller systems meet in groups. The gravitational forces exchange energies between different parts of colliding systems. As a result, a fraction of mass originally bound in smaller galaxies gets positive energy and leaves the system, while the remaining parts become more gravitationally bound. The star formation is enhanced during the merging events: a galaxy collision brings fresh gas to the galaxy central parts, where it triggers star formation.

While the main part of mass in disks of spiral galaxies is accreted in the form of cold gas, the majority of DM in clusters is acquired from mergers between DM galaxy haloes. A fraction of the primordial gas, and the gas that is released from galaxies in merging events, in gas stripping, and due to star formation feedback, resides in the intergalactic space of galaxy clusters forming intracluster medium (ICM).

A part of the galactic evolution is due to internal processes such as formation of spiral arms in rotating disks and evolution of bars in galaxy central regions. As an example of a barred spiral galaxy we refer to NGC 1365. It has two distinct spiral arms and the large central bar. The formation of bars and spirals will be discussed as a result of internal secular evolution of galaxy disks.

Star formation is another process of internal galaxy evolution. It produces individual stars, binaries or multiple stellar systems, it propagates in a turbulent ISM, where it is governed by different triggering mechanisms. Star formation transforms molecular clouds to expanding OB associations, or to more stable open star clusters. The original cloud is dispersed due to action of the star formation feedback: young stars produce intensive stellar winds and UV radiation penetrating the parent cloud. Formation of massive, stable and long-living stellar systems such as globular clusters needs some mechanism able to eliminate the dispersive action of the star formation feedback. Galaxy interaction may play a role in massive star cluster formation, since massive super-star clusters, which are the young progenitors of globular clusters, form preferentially in tidal arms or in the overlapping regions created when galaxies collide.

In this review, I will first focus on formation of stars and star clusters in molecular clouds, later I will describe evolution of galaxies due to gas accretion, mergers and secular evolution, then I will describe gas accretion and galaxy interactions near to the Milky Way, and finally I will discuss the interaction of individual galaxies with ICM in clusters. This provides only an incomplete picture of the galaxy evolution, some more and complementary information is provided by Palouš (2005).

## 2. STAR FORMATION IN THE ISM

Star formation is an important process of the internal galactic evolution. Stars are formed in cold, dense cores of molecular clouds. However, star formation may be triggered due to internal reasons, such as the star formation feedback, or due to external reasons, such as the galaxy collisions. We shall first discuss the star formation on its small scale inside of molecular clouds, later we shall mention different actions of triggering producing star formation patterns on larger scales.

Elmegreen (2007) describes the collapse and evolution of molecular clouds, which are hierarchical structures resulting from turbulence and self-gravity. The evolution of a molecular cloud has two different aspects: a fast collapse of its densest parts into stars, binaries, or small stellar groups, and a slow and long lasting dispersal of their low-density envelopes. There, in the envelopes, the next stellar generation may form due to triggering by energy and mass inserted by previous stellar generations.

The detailed SPH simulations of the collapse inside a molecular cloud are presented by Stamatellos et al. (2007): they consider the radiative transfer, which has an important influence on sizes of the resulting fragments. Kitsionas and Whitworth (2007) consider the case of a collision between dense clumps. The protostellar objects condense from filaments; they accrete their mass out of them. The effect of ionization due to radiation emitted by young stars is discussed by Dale et al. (2007) and Mac Low et al. (2007). The expanding HII regions collect mass in post-shock layers, where the star formation may be triggered. This collect and collapse model creates numerous self-gravitating fragments, particularly if the surrounding cloud is turbulent. The fragmentation times in their simulations correspond to theoretical predictions from the analysis of the expanding shell by Elmegreen (1994) and Whitworth et al. (1994). This proves that the dominant processes are mass accretion and stretching of isothermal, cold thin layers.

Stars emerge always in small filamentary groups, which later merge into star clusters. There is mounting evidence on non-coeval star formation in clusters. The structure of a cloud and its mass segregation is governed by the initial morphology producing the "primordial" mass segregation, rather than resulting from later relaxations (Sharma et al. 2007; Chen et al. 2007). What could be the origin of this early mass fragmentation? Whitworth et al. (1994) argued that high-mass stars form preferentially in compressed gas layers. This could lead to an opposite conclusion, since there should be more triggering by expanding shells on the periphery compared to the cloud center. The fragmentation to smaller parts in higher density places of an isothermal cloud, compared to lower density places, leads also to a conclusion opposite to observation. The explanation can be in the difference between the 3D collapse in the cloud center, where the density is high, but still less compared to shells, and the 2D collapse in thin cold shells, where the collect and collapse principle works: there, in shells, is the volume density higher compared to the cloud center, which is why the fragmentation process produces smaller fragments in shells at the cloud periphery compared to the 3D collapse in the cloud central parts.

Formation of massive stellar clusters, globular clusters and super star-clusters, which are composed of a million stars or more, is a challenging problem in astrophysics. To save the parental cloud from being dispersed, the star formation feedback inserting energy and mass into the star forming clouds needs to be stopped. It should increase the star formation efficiency to a high value, since the molecular cloud has to be transformed into stars almost completely.

Tenorio-Tagle et al. (2003) proposed the following scenario: the winds and radiation of the first generation of stars in the molecular cloud center creates an expanding shell, which propagates into the rest of the collapsing molecular cloud. At certain distance from the cloud center the shell stops expanding. It gradually accumulates the outer parts of the collapsing molecular cloud; it fragments producing stars of the super star-cluster.

The hydrodynamics of the stellar wind released by young stars into the volume of a super star-cluster is discussed by Tenorio-Tagle et al. (2007) and Wunsch et al. (2007). For given cluster radius, above threshold luminosity, the wind is thermally unstable. It produces dense and cold clumps, which may be the places of further star formation. This increases the fraction of the original cloud that is transformed to stars. The emerging super star-cluster is stable; it may become the progenitor of a globular cluster.

Why galaxy collisions trigger the formation of super star-clusters along the tidal arms? Molecular clouds are marginally stable: in an environment with certain pressure, their self-gravity is in a balance with random motions and galactic tides. A galaxy collision triggers their collapse due to increased external pressure in the molecular cloud environment (Jog & Solomon, 1992). Another triggering factor may be the decrease of tidal fields along tidal arms created by the interaction (Palouš et al. 2004). This hypothesis should be checked with velocity fields of interacting galaxies such as NGC 6621/2 or NGC 5752/4.

### 3. GAS ACCRETION, MERGERS AND SECULAR EVOLUTION OF GALAXIES

The evolution of galaxies is a combination of external processes, such as cold gas accretion from intergalactic spaces, or hierarchical merging of small irregular and disk galaxies in clusters, with internal processes, such as the galaxy secular evolution redistributing mass and angular momentum in galaxies, or star formation. Cold gas accretion, which in the $\Lambda$CDM model of the expanding universe (Gnedin 2003) occurs preferentially along the large scale filaments of the mass distribution, is able to build up the galaxy disks. They are the subjects of later secular evolution. Hierarchical merging of small irregular and disk galaxies in groups and clusters produce larger elliptical galaxies, where the strongly rotating disks are absent. Merging applies particularly to collisionless dark matter, when it combines and builds the gravitational potential of a galaxy cluster. Secular evolution of the disks forms bars and spiral arms, and gradually builds up central mass concentrations (CMCs).

The galaxy growth by accretion is described by Semelin and Combes (2005). According to their simulations, accretion dominates the galaxy growth by a factor 2 to 4 over the mass gained through mergers. In the case of field galaxies is this fraction even higher. The dominance of the accretion applies more to $z > 2$ and decreases from 4 times to 2 times between $z = 2$ and $z = 0$. In the local volume at $z = 0$ is the accretion more anisotropic with an excess along the plane of the galaxy disk. There is an observational evidence of cold gas accretion on galaxies as shown by van der Hulst and Sancisi (2004). The nearby case of gas accretion to the Milky Way, the Magellanic Stream, was observed in HI by Brüns et al. (2005). It will be discussed in more details later.

The hierarchical theory of galaxy formation assumes that small irregular galaxies form first, later they merge forming larger galaxies. Bournaud et al. (2005) concluded that the mass ratio of the collision partners is the most important parameter determining the global properties of the remnant. The major mergers with the mass ratio in the range 1:1 – 4:1 produce remnants with the density profile and kinematics similar to elliptical galaxies, the intermediate mergers in the range 4:1 – 10:1 form remnants that could be progenitors of S0 galaxies, and minor mergers with mass ratio more than 10:1 result in disturbed spiral galaxies, with the thick disk components.

The bars and spiral arms in disk galaxies result from their secular evolution. Debattista et al. (2006) show that the disk-like properties of the galaxy bulges and the links between bulge and disk properties indicate that the bulges are formed through the evolution of the disk. Bulges that often have box- or peanut-shapes are associated to the evolution driven by a central bar. The secular evolution changes the disk profile that does not resemble its original shape. It is an open task to disentangle what properties of the disk result from its internal evolution and what result from hierarchical or gas accretion assembly.

The early numerical $N$-body experiments with rotating disks by Ostriker and Peebles (1974) did show that, when the ratio $Q_\mathrm{B}$ of the total kinetic energy in organized rotational motion $E_\mathrm{rot}$ to the total gravitational energy $E_\mathrm{g}$:

$$Q_\mathrm{B} = E_\mathrm{rot} / E_\mathrm{g} > 0.14 \,,$$

the bar forms in the central part of the disk. It is supported by important families of stellar orbits making it rather stable configuration.

In the outer parts of the disk, the evolution depends on the fraction of the energy in random motions relative to gravitational forces. In a flat disk, this ratio is given with a parameter $Q$ derived by Toomre (1964, 1977):

$$Q = \sigma_\mathrm{R} \kappa / (3.36 \, G \Sigma) \,,$$

Where $\sigma_\mathrm{R}$ is the radial component of the velocity dispersion in the disk, $\kappa$ is the epicyclic frequency, $\Sigma$ is the disk surface density and $G$ is the constant of gravity. When $\sigma_\mathrm{R}$ is small enough, the disk is cold, and consequently $Q$ is small. In this case it is unstable to perturbations, and it forms spiral arms. The gravitational field of spiral arms perturbs the stellar orbits and increase $\sigma_R$ increasing $Q$, which stabilizes the disk.

The spiral galaxy disks include besides stars also the ISM. This is a dissipative component able to decrease its random motions due to inelastic collisions between the ISM clouds or in supersonic turbulent motions. It agitates spiral arm formation in the ISM, where the spirals are more pronounced compared to stars.

The gas flow lines are systematically shifted with respect to the stellar orbits, forming a gravity torque. This gravity torque in combination with a viscosity torque, which is connected to the inelastic collisions between the ISM clouds, are the reasons for the gradual mass concentration to the galaxy central part forming CMC. Stellar orbits scatter on the CMC, which weakens the central bar. Bournaud et al. (2005) analyzed the lifetime of galactic bars: they weaken due to growing CMC, and due to the gravity torque and angular momentum transfer from the infalling gas to the stellar bar. Their lifetime is finite; according to estimates of Bournaud et al. (2005) the bar exists 1–2 Gyr.

The above model of the secular evolution of galaxies is summarized by Combes (2007): the bar forms in a disk, when it accumulates sufficient cold gas from accretion or minor mergers. The bar and spiral arms create a gravity torque, which causes the inflow of the gas towards the center. The inflowing mass builds the CMC. Scattering of star on the CMC and the gravity torque between the gas flow lines and stellar orbits restrict the lifetime of the bar to 1–2 Gyr. A dissolved bar turns to the central galactic bulge. The reformation of bars needs new accretion of cold gas from outside the galaxy.

## 4. LMC, SMC AND THE MILKY WAY SYSTEM

A nearby example of the gas accretion is the interacting system Milky Way, Large and Small Magellanic Clouds (MW–LMC–SMC). The HI observations by Brüns et al. (2005) showed large scale features such as Magellanic Stream, Magellanic Bridge, Interface Region and the Leading Arm. These features are the result of the gravitational interaction between the three galaxies, maybe in a combination with the gas stripping due to diluted hot gas in the galactic halo. The role of stripping is still not clear, since the plausible model of the interaction explaining the formation of the observed features is not accepted. However, in this case, the cold gas is accreted to the MW disk periphery. There is no direct influence on the disk inner parts, where the MW bar resides.

An investigation of test particle orbits in the galactic halo allows mapping extended parameter spaces with the genetic algorithm using a fitness function to compare the observed features to simulations (Růžička et al. 2007). This search shows that if the DM halo of the MW would be flattened the observed features are slightly better reproduced compared to spherical or prolate halo shapes.

The output is sensitive the transversal velocities of the Clouds. The new estimates based on HST proper motions (Kallivayalil et al. 2006) discovered that the transversal velocities of the Clouds are substantially larger than previously estimated from HIPPACOS proper motions (Kroupa and Bastian 1997). If correct, and with the same total mass of the Milky Way, it may suggest that the Magellanic Clouds are on their first passage around the Milky Way (Besla et al. 2007). However, the first passage model has to reproduce all the morphological features, and it is not clear if there are some orbits tuning the tidal forces in such a way that the observation is fitted. There may also be other constraints, such as total mass of the MW–LMC–SMC system, or initial gas distributions in the galaxy disks, necessary to reproduce the large-scale shapes of the cold gas features.

### 5. GAS STRIPPING IN GALAXY CLUSTERS

The accretion of cold gas and merger events bring mass to galaxies, contributing to their growth. These processes supply fresh gas used for star formation, increasing the star formation rates. The feedback of energy and mass from young stars can decrease the mass inflow or even revert it in some places to outflows. Star formation feedback provides metals to the ISM, and it may eject the yields to the ICM, which is a large depository of metals (Renzini 2004).

Butcher and Oemler (1978, 1984) compared distant clusters of galaxies to nearby clusters: they concluded that there is a higher fraction of blue star forming galaxies in distant relative to nearby clusters. This substantial evolution in clusters may be due to several effects. One of them is the gravitational perturbation when a galaxy meets along its orbit another one. The encounters at velocities of a few 1000 km/s are described as galaxy harassment events by Moore et al. (1995): they influence the structure of galaxies; multiple harassment events can remove galaxy peripheries and change a spiral into an S0 type. Other effects are the tides from the cluster potential itself, or hydrodynamic influence of the ICM called ram pressure gas stripping.

The motion of individual galaxies in the ICM of galaxy clusters exerts the ram pressure on the ISM. Its effect has been predicted by Gunn and Gott (1972), who assumed that after the formation of a galaxy cluster, the remaining gaseous debris are thermalized via shock heating to temperatures above 10 million K. This diluted gas of low density influences the ISM of spiral galaxies in clusters due to ram pressure. The ISM is stripped from galaxies, which also changes the star formation: it is increased at the leading edge of the galaxy motion, and it is quenched in other places.

The effect of gas stripping is discussed by Jáchym et al. (2007) in numerical simulations with the code GATGET using the SPH with a gravity tree. The role of galactic orbital history on the stripping amount is stressed. Pure stripping may be responsible for gas accumulation in tails, filaments and ripples in regions with low overall ICM occurrence. However, the total amount of the ICM in clusters cannot be produced by pure ISM stripping due to ram pressure; it needs to be combined with tides and stellar feedback. A fraction of the ICM has to have "primordial" origin.

*Acknowledgments.* I would like to thank the organizers of the astrophysical school "Heliosphere and Galaxy" and in particular Dr. Nedelia Popescu for her invitation to beautiful Sinaia. I also like to express my thanks to Bruce G. Elmegreen for his comments on an early version of the paper. This review has been supported by the Institutional Research Plan AV0Z10030501 of the Academy of Sciences of the Czech Republic and by the project LC06014 Center for Theoretical Astrophysics.